\documentclass[twocolumn]{elsart}

\usepackage{graphicx}

\usepackage{natbib}

\newcommand{\be}{\begin{equation}}
\newcommand{\ee}{\end{equation}}
\newcommand{\nn}{\mbox{} \nonumber \\ \mbox{} }
\newcommand{\ba}{\begin{eqnarray}}
\newcommand{\ea}{\end{eqnarray}}

\newcommand\eg{\textit{e.g.\ }}
\newcommand\cf{\textit{cf.\ }}

\newcommand\lo{\mathrel{\raise.3ex\hbox{$<$}\mkern-14mu\lower0.6ex\hbox{$\sim$}}}
\newcommand\go{\mathrel{\raise.3ex\hbox{$>$}\mkern-14mu\lower0.6ex\hbox{$\sim$}}}

\newcommand\apjl{ApJ Lett.}
\newcommand\apj{ApJ}
\newcommand\mnras{MNRAS}
\newcommand\jgr{Jour. Geophys. Res.}

\begin{document}
\runauthor{Lyutikov}
\begin{frontmatter}
\title{Eclipses and orbital modulations
in binary pulsar  PSR  J0737$-$3039}
\author{M. Lyutikov}
\address{University of British Columbia, 6224 Agricultural Road,
Vancouver, BC, V6T 1Z1, Canada}

\begin{abstract}
Binary  radio pulsar system J0737$-$3039 provides an exceptional opportunity to 
study innermost structure of pulsar magnetospheres due to 
very tight orbit,  favorable  directions of pulsars' rotation and magnetic axes
and extremely 
fortuitous 
orientation of the orbit.
In this system  the millisecond pulsar A is 
eclipsed once per orbit. During eclipse a 
clear modulation  at the 2.77 s 
period of pulsar B 
is seen, pointing unambiguously to magnetospheric origin of eclipses.
A simple geometric model, based
on the idea that the radio pulses are attenuated by synchrotron
absorption on the
closed magnetic field lines of 
pulsar B, can successfully reproduces the eclipse light curves down to intricate details.
This detailed agreement confirms the 
 dipolar structure of the neutron star's  magnetic field.
 The model gives clear predictions for temporal evolution of eclipse profile
 due to geodetic precession of pulsar B.

In addition,  pulsar B
shows orbital  modulations of intensity, being especially bright at two
short orbital  phases. We showed that  these modulations are due to
distortion of pulsar B magnetosphere
by pulsar A wind which produces orbital phase-dependent changes
of the    direction
along which radio waves are emitted. Thus, pulsar   B is intrinsically
bright at all times but its radiation  beam misses the Earth
at most  orbital phases.
\end{abstract}

\end{frontmatter}

\section{Amazing and lucky: double pulsar  PSR J0737$-$3039A/B}

\begin{figure}[h]
   \includegraphics[width=0.95\linewidth]{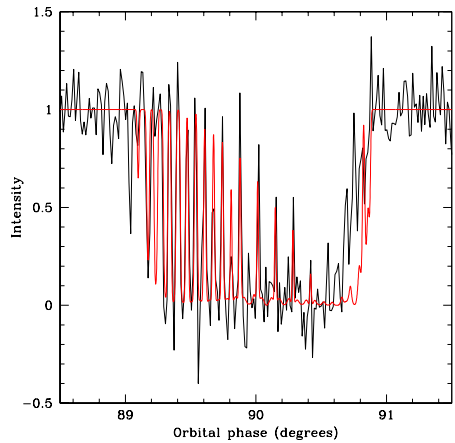}
   \caption{
    Comparison of a simulated eclipse profile  with 800 MHz data
    \citep{mcla04}. 
    The model fits the data best in the middle of the eclipse.
 }
 \label{compare}
 \end{figure}

The double pulsar system PSR J0737$-$3039A/B contains a recycled 22.7 
ms  pulsar (A) in a 2.4 hr orbit around a 2.77 s pulsar (B)  
\citep{burgay03,lyne}. The system is viewed nearly edge on,
with the line of sight making an angle $\sim 0.5^\circ$ with the orbital plane
\citep{coles,ransom04}.
This leads to a number of exceptional properties of the system.
Pulsar A is eclipsed once per orbit, 
for a duration of $\sim 30$ s 
centered around superior conjunction (when pulsar B is between the observer and pulsar A).
The width of the eclipse is only a weak function of the observing frequency
\citep{kaspi}. Most  surprisingly, during eclipse the pulsar A radio flux is
modulated by the rotation of pulsar B: 
there are  narrow, transparent windows in  which
the flux from pulsar A rises nearly to the unabsorbed level \citep{mcla04}.
These spikes in the radio flux are tied to the
rotational phase of pulsar B, and provide key constraints on the geometry of the
absorbing plasma. At ingress
transparent spikes appear at half rotational period of B, then change their
modulation to rotational period of B at the middle of eclipse and disappear completely
right before the egress (Fig \ref{compare}).
On average
the eclipse is longer when the magnetic axis of
pulsar B is approximately aligned with the 
line of sight (assuming that radio pulses are generated near the
magnetic axis of B).

The physical width of the region which causes this periodic modulation is
comparable to, or somewhat smaller, than the estimated size of the magnetosphere
of pulsar B.  Combined with the rotational modulation, 
this provides a strong hint that the absorption is occurring
{\it within} the magnetosphere of pulsar B.  This allows us to probe directly 
the structure of pulsar  B magnetosphere.  

We constructed a detailed model of eclipses \citep{lt05}
which shows that the light curve is consistent, in considerable detail, with 
synchrotron absorption in  magnetosphere of pulsar B.
The agreement is detailed enough to constitute direct evidence 
for the presence of a {\it dipolar} magnetic field around pulsar B.   

\section{Magnetosphere of pulsar  B}

At distances much larger than neutron star radius but 
much smaller than light cylinder radius, 
$R_{LC} \sim c/\Omega $ ($\Omega$ is rotation frequency of a
pulsar,
magnetospheres of {\it isolated} pulsars 
are nearly dipolar, containing two types of magnetic field lines: 
those that close inside $R_{LC}$
and those that cross 
the light cylinder and thus remain open to infinity \citep{gj69}. 
Radio emission is, presumably, generated on the open field lines near the neutron star
magnetic polar field line.

In case of PSR J0737$-$3039B this picture is modified by strong interaction
of pulsar B with relativistic wind flowing outward from pulsar A.
Wind of A strongly distorts B  magnetosphere, blowing off  a large 
fraction of it, so that 
  the size of magnetosphere
 of B facing A 
 is much smaller than light cylinder and is now determined by 
 the pressure balance
between the wind of pulsar A and the magnetic pressure of pulsar B 
\cite[\eg][]{lyut}.
Observationally, this is confirmed by the fact that the half-width of the eclipse,
 $\sim  1.6\times 10^9$ cm, is much smaller than the pulsar B light cylinder $R_{LC} \sim 1.3 \times 10^{10}   $ cm.

As a results of interaction with A wind, 
 spindown torque on pulsar B is  modified if compared with the vacuum case.
This has important implications for pulsar B  magnetic field and spindown age, which
 can
be estimated self-consistently only by modeling the interaction
between wind and magnetosphere (conventional vacuum  spindown formulas are 
incorrect in this case!).
Estimates of the  torque \citep{lyut,arons} show
that {\it breaking index  of pulsar B is likely to be close to $1$}
and may even approach $0$ 
(instead of vacuum dipole case of $3$). Surface magnetic field 
and magnetospheric radius can be estimated as
\ba &&
B_{\rm NS}  
\simeq {D_{AB}^{1/2} \left( I_B  \dot{\Omega}_B/\Omega_B \right)^{3/4} c
     \over  \left( I_A \Omega_A \dot{\Omega}_A \right)^{1/4} R_{\rm NS}^3  }=
 \nn &&
 4 \times 10^{11} \,{I_{B,45}^{3/4}\over I_{A,45}^{1/4}}
    \;\;\;\; {\rm G};
\nn &&
 R_{\rm mag} \simeq  \left[ { (cD_{AB})^2 I_B  \dot{\Omega}_B \over 
 I_A \Omega_A \dot{\Omega}_A \Omega_B} \right]^{1/4}=
\nn &&
4.0 \times 10^9\, 
\left({I_B\over I_A}\right)^{1/4}\;\;\;\; {\rm cm}
\label{Bn}
\ea
where $ \Omega_{A,B}$ and $\dot{\Omega}_{A,B} $ are measured spin frequencies and 
frequency derivatives \citep{burgay03},
 $I_{A,B}$ are moments of inertia normalized to $10^{45}$ g cm$^2$ and
 $D_{AB}$ is distance between pulsars.

As long as parameters of pulsar A remain approximately constant, over time scale of 250  Myrs,
pulsar B spins down exponentially, $\Omega_B \propto \exp\{- t/\tau\}$
where $\tau = { \Omega_B  / \dot{ \Omega}_B} \approx 100 $  Myrs 
(note that this is two times larger than its characteristic age).

While details of  magnetosphere-wind interaction are bound to be complicated,
a reasonable guess, on which the above 
 estimates  are  based, is that  
open-field current density remains 
similar to the one in isolated pulsars, $j \sim n_{GJ} e c$
($n_{GJ} = \Omega B / 2\pi e c$ is Goldreich-Julian density), 
while the size of open field lines  is increased due to
smaller size of B magnetosphere.
In fact, field dragging at the magnetospheric boundary may 
increase typical current density 
deep 
inside magnetospheres. 
This will lead to increased torque and decreased estimates
of surface magnetic field and  magnetospheric radius. 
Still, we can estimate 
the minimum magnetospheric radius by assuming that at the boundary toroidal magnetic field
does not exceed poloidal magnetic field. This gives a minimum estimate
of the magnetospheric radius (corresponding to spindown index $0$):
$
R_{\rm mag, min } =
\left({  I_B  \dot{\Omega}_B
\over I_A \Omega_A \dot{\Omega}_A}   \right)^{1/3}
  \left({ c D_{AB}^2 \over  2  } \right)^{1/3} =2.4 \times 10^{9} {\rm cm}
$.
Note, that this is still somewhat larger than the size  of eclipses.

\subsection{Model of eclipses}

Similar to the case of isolated pulsars, we assume that  magnetic fields lines of pulsar B
are of two types: open and closed ones. In addition, we assume that 
closed magnetic fields lines
extend not up to the
light cylinder
 but up to the magnetosphere radius $R_{\rm mag}$, while field lines 
extending further than $R_{\rm mag}$   remain open.
We place 
 the 
absorbing relativistically hot 
plasma  within a set of closed dipolar  field lines.
It can be shown that plasma density is constant along each field line.
We further assume that density and temperature do not vary {\it between } field lines.
Both of these assumptions have only minor effect on eclipse profile.

We 
calculate the synchrotron optical depth over a large number of lines of sight,
taking into account both the three-dimensional structure
and rotation of B magnetosphere
(Fig. \ref{movie}).
The calculation of the eclipse light curve requires a choice of
several parameters:  direction of rotational axis (angles $\theta_\Omega$, $\phi_\Omega$), 
angle between rotational and magnetic axis
$\chi$, impact parameter $z_0$, 
electron temperature $T_e$, density multiplicity $\lambda_{\rm mag}$ (ratio of density
to Goldreich-Julian density)
and outer radii $R_{\rm abs+}$

Results of the simulations are compared with data on in Fig. (\ref{compare})
 {\it  The model  reproduces many fine details of the
 eclipses.}
The model can reproduce asymmetric form of 
eclipse with the ingress shallower and longer than
the egress. 
It explains the modulation of that is observed at the first
and second harmonics of the spin frequency of pulsar B, and the deepening
of the eclipse after superior conjunction.
The average eclipse duration is almost independent of frequency
when  multiplicity is sufficiently  large.
The onset and termination of the eclipse are determined mostly
by the physical boundary of the absorbing plasma and not by   microphysics
of absorption process. This results in nearly frequency independent eclipses.
 Eclipse is  broader  when
the magnetic moment of pulsar B is pointing closest to the observer,
just as is observed \citep{mcla04}.  
There are modest deviations between the model and the data near the edges of
the eclipse that  could be used to probe the distortion of magnetic field
lines from a true dipole.

The modulation of the radio flux during the eclipse is due to 
the fact that -- at some rotational phases of pulsar B -- the line 
of sight  only passes through open magnetic field lines where absorption
is assumed to be negligible.
One of the main successes of the model is its ability
to reproduce both the single and double periodicities of these
transparent windows, at appropriate places in the eclipse.  
This requires that $\vec\mu_B$ be approximately -- but not
quite -- orthogonal to ${\bf\Omega}_B$.

Based on a sample of many eclipse light curves 
our best fit  parameters are:   
$z_0 \simeq -7.5\times 10^8$ cm,  
$\theta_\Omega\simeq 60^\circ $,
$\phi_\Omega \simeq -90 ^\circ$,  $\chi \simeq 75^\circ$.
The orbital inclination is therefore predicted to be close to
$90.55^\circ$, \cite[\cf][]{coles,ransom04}.
The size of the eclipsing region is $R_{\rm abs+} =1.5 \times 10^9$;
the  implied
multiplicity is large, $\lambda_{\rm mag}= n_{mag}/n_{GJ} \sim 10^5$. Lorentz factor of suspended plasma was 
fixed at $\gamma=10$ (there is a degeneracy between Lorentz factor and density).  

Presence of  high multiplicity,
relativistically hot plasma on closed field lines of pulsar B is somewhat surprising, but not unreasonable. 
Dense, relativistically hot plasma can be effectively stored
in the outer magnetosphere, where cyclotron cooling is slow.  
The gradual loss of particles inward through the cooling radius, occurring on time
scale of millions of pulsar B periods, can be easily
compensated by a relatively weak  upward flux driven by a fluctuating component
of the current. For example, if suspended material is resupplied at a rate of 
one Goldreich-Julina density per B period and particle residence time is million
periods, equilibrium density will be as high
as $10^6 n_{GJ}$. 
The trapped particles  are also heated to relativistic
energies by the damping of magnetospheric turbulence \citep{lt05}.

There is only one slight deviation from simple estimates:
our eclipse calculations show that the optical depth of the 
absorbing plasma undergoes a sharp drop at a distance  $R_{\rm abs+} \simeq
1.5\times 10^9$ cm from pulsar B.  
This is about $2.5$ times smaller than the expected radius
of the magnetopause, $R_{\rm mag} \simeq 4\times 10^{10}$ cm.
Most likely  this is due to 
 the loss of plasma from the 
outermost closed field lines due to reconnection and/or gradient drift.

\subsection{Predictions of the model}
There is a number of predictions of the model that should be tested in the coming years.
\begin{enumerate}
\item
 The spin ${\bf\Omega}_B$ of pulsar B is expected to undergo
geodetic precession on a $\sim 75$ year timescale \citep{burgay03}.
We have provided predictions for how the eclipse light curve will
vary as a result.  In particular, the orbital phase at which the
radio flux reaches a minimum will shift back toward superior conjunction.
 Since $\phi_\Omega$ is not well constrained, 
a time for eclipse to become symmetric is between $\sim$ 12 years (if 
$|\phi_\Omega+90^\circ| \sim 30^\circ$ and ${\bf\Omega}_B$ is 
drifting  away from the plane of the sky) 
and $\sim$ 25 years (if $\phi_\Omega$ is drifting toward the plane of the sky).

\item
High time-resolution observations are a sensitive probe
of the distribution of plasma properties (density and temperature) on the
closed field lines. If plasma is depleted from  the outermost
field lines, at high temporal resolution the flux should return to unity.
On the other hand, if the absorbing plasma does not
have a sharp truncation in radius, then
the flux will not return to unity in all of the transparent windows.

\item
The eclipses must regain 
a strong frequency dependence at sufficiently high frequencies.  The critical
frequency above which significant transmission occurs
can be used to place tight constraints on the plasma density.
The electron cyclotron frequency at a distance $r \sim z_0 \sim 7.5\times 10^8$
cm from pulsar B is estimated to be $\nu_{B,e} \sim 3$ GHz. 
One therefore expects the eclipses to develop a significant frequency
dependence at higher frequencies.

 \item There are a number of
 definite predictions for the polarization of the transmitted
radiation \citep{lt05}. But since pulsar A emission is strongly  polarized one needs to 
know absolute position of the direction of linear polarization.

\end{enumerate}

\section{Orbital modulation of B due to 
distortion of B magnetosphere by A wind}

Next we turn to  orbital  modulations of intensity of J0737$-$3039B. 
Emission of pulsar B is strongly dependent on  orbital phase, being 
 especially bright at two windows,  each lasting for about $30^\circ$
 \citep{lyne}.
Pulse profiles have different shapes in the two windows.
We showed \citep{lyut05}   that  orbital brighting of B is not intrinsic
but is due to distortions of  B magnetosphere by pulsar A wind, so that
 radiating beam of B mostly misses the observer, except at two orbital phases. 

To prove this point,
we  assume that intrinsically bright pulsar B radio  emission is generated near 
the polar field line inside a region
with half opening  angle of $\sim
2 $ degrees, while the polar field line itself 
is ''pushed around'' by the pressure of A wind. 
 To parametrize the effects of A wind on direction of  radio emission of B 
  we employ a  method of distortion transformation of
  Euler potentials \citep{Stern94},
  developed for modeling of the Earth  magnetic field
   influenced by solar wind.
Assuming that radio emission of B is generated in physically small
region of B magnetosphere (much smaller than its size), 
direction of  radio emission of B
can be parametrized by a distortion coefficient $C$ \citep{lyut05}
(so that distortion is proportional to $1-C$, for pure dipole $C=1$).

\begin{figure}[h]
\includegraphics[width=0.9\linewidth]{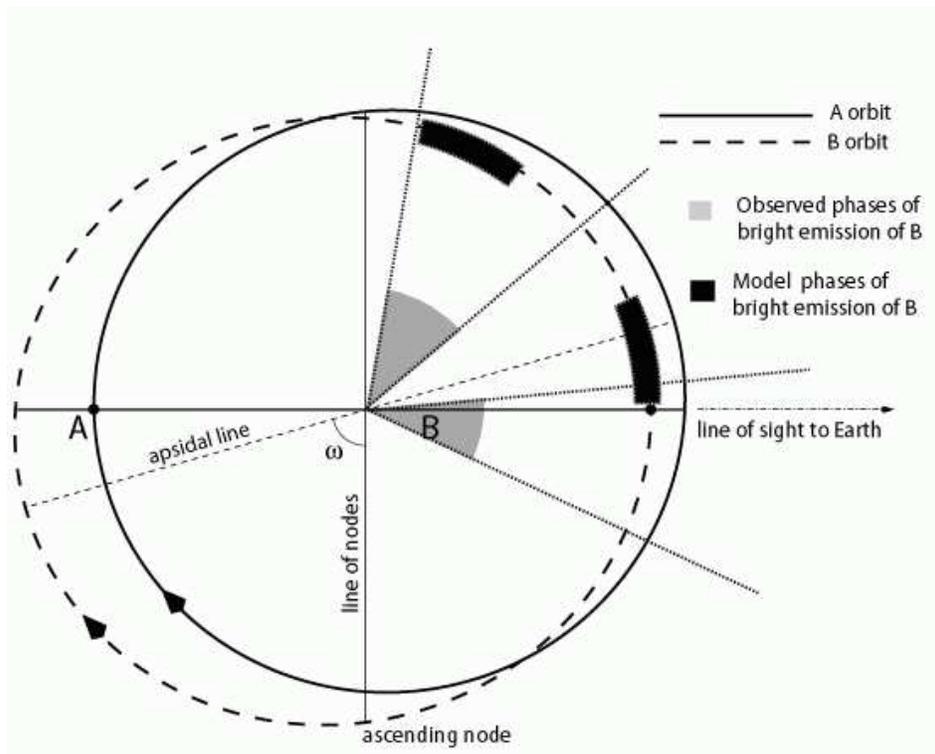}
\caption{Configuration of the system, after \protect\cite{lyne}. Light
shades segments indicate
 orbital phases where B emission is strongest, dark  regions  indicate  the
 location of emission in our best fit model.
 }
 \label{emissionphase}
 \end{figure}

To fit the data  we use the results of
modeling of pulsar A eclipse
\citep{lt05}. Searching through parameter space we were 
able to obtains a satisfactory fit, producing
emission at two orbital phases nearly coincident with observed ones,
Fig. \ref{emissionphase}.

The strength of the model is that it reproduces (or predicted)
fairly  well a number of properties of pulsar B radio emission.
First, 
since at different orbital phases the
line of sight makes different cuts through emission region,
the model naturally explains orbital phase-dependent averaged emission profile.
In fact, this can be used to construct a map of the emission region, something 
which was not possible for isolated pulsars.
Secondly, 
since the emitting beam of B is pointing in somewhat different directions at 
different orbital phases, time of arrival of  
pulses of B should experience  regular orbital-dependent changes  by as much as
$\sim (2^\circ/360^\circ) 2.8 {\rm sec} \sim 10 $ ms.
 Such  drift in times of arrival has  been observed:  \cite{ransom04}  reported a
 systematic change in arrival times of B pulses by 10-20 ms.

Finally, since  emission beam of B  is fairly narrow, small changes  in the
orientation of rotation axis of B may induce large apparent changes
in the profile. One possibility for the change in the direction
of the spin is  geodetic precession of B, which should happen
on a relatively short time scale $\sim 70 $ yrs \citep{lyne}.  From our modeling
we find that changes of
$\phi_\Omega$ by as little as $\sim 1^\circ$ strongly affect observed B
profile.  Thus, we expect that profile of B may change on a times scale of less than a  year. 
 \footnote{According to the model,
  we are not likely to lose pulsar B in the coming year(s), yet the model
   is not sufficiently detailed to guarantee it.}
This prediction has been recently  confirmed by \cite{burgay05}, 
who see  evolution of profile of 
PSR J0737$-$3039B in general agreement with our model.

Another possible implication of the model
concerns emission heights in pulsar B.  Data are best fitted by the model with
the stretching coefficient $C=0.7$, so  
relative deformation of the magnetosphere at the emission radius is $\sim 30\%$.
This is a fairly large distortion, which
 favors  large emission altitudes,
 $R_{em}\geq  10^8$ cm. Large emission altitudes in isolated pulsars
 have been previously suggested by  \cite{lbm98}.

The main implication of the model is that B is always intrinsically bright.
Its peak luminosity of $\sim 3$ mJy at 820MHz is, in fact,  typical for isolated pulsars with similar properties.
There is a number of ways the model can be improved. First, a better model of 
dipole distortion is needed. This can be achieved by using well developed models of the
Earth magnetosphere: since the size of pulsar B  magnetosphere is much smaller than light cylinder, light travel effects are probably unimportant.
Second, a non-trivial
geometry of the emission region, as opposed to a circular used in the present study,
may increase
 the quality of the fit.  
Also, if reconnection between
wind and magnetospheric field lines is important, the structure of  magnetosphere may depend
on the direction of the wind magnetic field.
We plan to address these issues in a subsequent paper.

\section{Conclusion}
\label{concl}

The success of the models  is somewhat surprising,
given the complexity of data   and the
fact that  we had to fit
many parameters.
The  reasons for it 
is that, to the first order,
magnetic field of B is  well approximated by dipolar structure.
Thus, the models provide a strong test of the longstanding assumption
that neutron stars are surrounded by corotating, {\it dipolar} magnetic
fields. This is a valuable confirmation of a fundamental assumption made in
models of pulsar electrodynamics.
In  addition, our results are consistent with
models that place the source of the radio emission close to the magnetic axis.
Forthcoming improvement of the model should be able to quantify
deviations of the magnetosphere from dipolar shape and  probe  how a pulsar magnetosphere interacts
with an external wind.

Our eclipse model  is consistent with a large natal kick of pulsar B that
changes the orientation of the orbital plane
and disrupt any pre-existing alignment between orbit and the
spin of the progenitor star.

{}

\begin{onecolumn}

\begin{figure}[h]
 \includegraphics[width=0.9\linewidth]{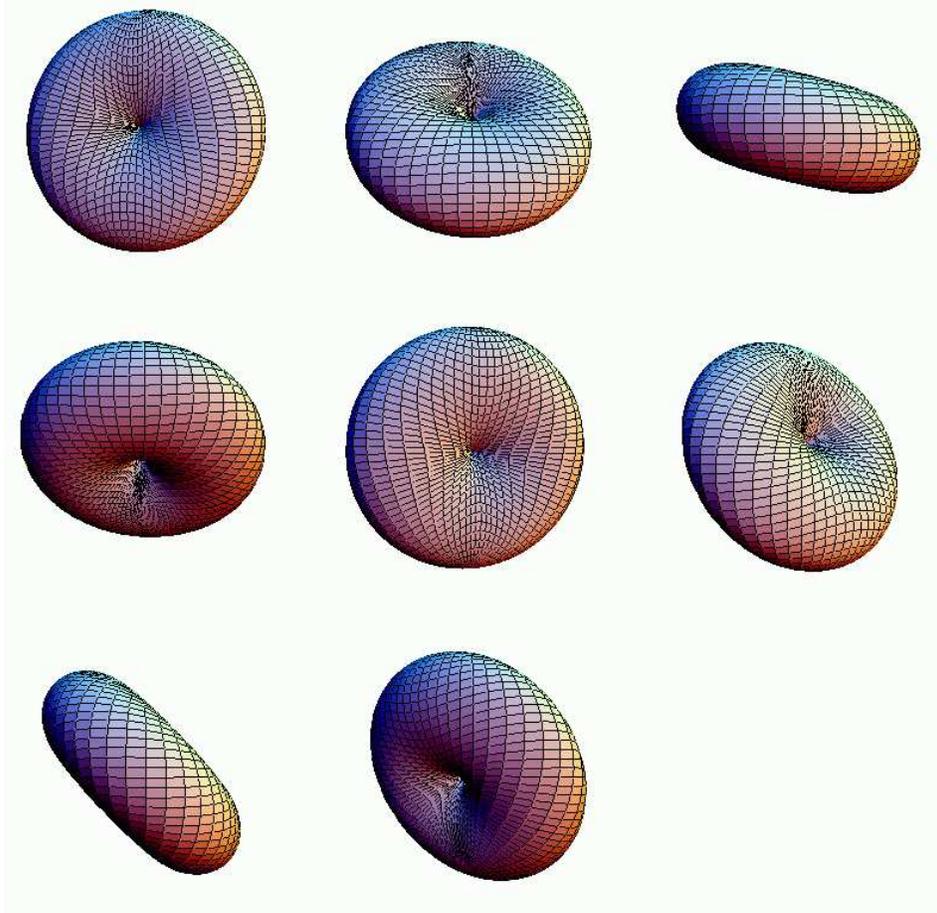}
\caption{View of the magnetosphere at different rotational phases
 separated by $\pi/4$.  For a full movie of the eclipse see 
 http://www.physics.mcgill.ca/$\sim$lyutikov/movie.gif}
  \label{movie}
  \end{figure}

\end{onecolumn}

\end{document}